\documentclass[superscriptaddress,preprintnumbers,amssymb,nofootinbib]{revtex4-2}
\usepackage[left]{lineno}
\usepackage{blindtext}
\pdfoutput=1
\usepackage[T1]{fontenc}
\usepackage{graphicx}
\usepackage{epsfig}
\usepackage{amsmath}
\usepackage{amsfonts}
\usepackage{amssymb}
\usepackage{braket}
\usepackage{cancel}
\usepackage{pstricks}
\usepackage{color}
\usepackage{feynmf}
\usepackage[normalem]{ulem}
\usepackage{epstopdf}
\usepackage{soul}

\definecolor{ao(english)}{rgb}{0.0, 0.5, 0.0}

\usepackage{subcaption}
\usepackage{braket}
\usepackage{cancel}
\usepackage{setspace}
\usepackage{pstricks}
\usepackage{xcolor}
\usepackage{float}
\usepackage{mathtools}
\usepackage{multirow}
\usepackage{booktabs}
\usepackage{enumitem}
\providecommand{\tabularnewline}{\\}

\def\missET {{\not\!\! E_T}}
\def\gsim{\lower0.5ex\hbox{$\:\buildrel >\over\sim\:$}}
\def\lsim{\lower0.5ex\hbox{$\:\buildrel <\over\sim\:$}}

\begin{document}


\title{High $p_T$ correlated tests of lepton universality in lepton(s) + jet(s) processes; an EFT analysis}

\author{Yoav Afik}
\email{yoavafik@campus.technion.ac.il}
\author{Shaouly Bar-Shalom}
\email{shaouly@physics.technion.ac.il}
\author{Jonathan Cohen}
\email{jcohen@campus.technion.ac.il}
\affiliation{Physics Department, Technion--Institute of Technology, Haifa 3200003, Israel}
\author{Amarjit Soni}
\email{adlersoni@gmail.com}
\affiliation{Physics Department, Brookhaven National Laboratory, Upton, NY 11973, US}
\author{Jose Wudka}
\email{jose.wudka@ucr.edu}
\affiliation{Physics Department, University of California, Riverside, CA 92521, USA}

\date{\today}

\begin{abstract}
We suggest a new class of tests for searching for lepton flavor non-universality (LFNU) using  ratio observables and based on correlations among the underlying LFNU new physics (NP) effects in several (seemingly independent) di-lepton and single lepton + jet(s) processes.
This is demonstrated by studying the effects generated by  LFNU 
4-Fermi interactions 
involving  3rd generation quarks.
We find that 
the sensitivity to the scale ($\Lambda$) of the LFNU
4-Fermi operators significantly improves when the correlations among the various di-lepton +jets and single-lepton + jets processes are used, reaching $\Lambda \sim {\cal O}(10)$~TeV at the HL-LHC. 
\end{abstract}

\maketitle
\flushbottom


Intriguing hints of lepton-flavor non-universality (LFNU) and therefore of new physics (NP) have appeared in recent years in neutral and charged semileptonic B-decays  \cite{Aaij:2014pli,Aaij:2014ora,Aaij:2017vbb,Aaij:2015esa,Aaij:2015oid,Wehle:2016yoi,Abdesselam:2016llu,ATLAS:2017dlm,CMS:2017ivg,Bifani:2017gyn,Aaij:2019wad,Abdesselam:2019wac,Lees:2012xj,Lees:2013uzd,Huschle:2015rga,Hirose:2016wfn,Aaij:2015yra,Aaij:2017uff,Aaij:2017deq,Adamczyk:2019wyt,Abdesselam:2019dgh} (for a recent review see \cite{Bifani:2018zmi}): the $R_{K^{(*)}}$ and $R_{D^{(*)}}$ anomalies which occur in $b\to s \ell^+ \ell^-$ and $b\to c \ell^- \nu_{\ell}$ transitions, respectively.

In this work, we consider testing for  LFNU in lepton(s) + jets production at the LHC, by exploiting correlations amongst several LFNU observables. Specifically, we show that an enhanced sensitivity to the scale of the NP can be obtained by combining multiple LFNU tests, based on ratio observables. We demonstrate this for two specific new physics (NP) scenarios, using the so-called SM  Effective Field Theory (SMEFT) framework \cite{EFT1,EFT2,EFT3,EFT4}, although this approach can be extended to establish a more systematic mapping between the underlying NP dynamics and experimentally realistic observables. 
The importance of using correlations in the search for NP has recently gained some attention, e.g., in leptoquark searches by combining di-lepton and single lepton production channels \cite{Bansal:2018eha} and
in top-quark systems by using measurements from different top production and decay processes 
to probe the NP effects \cite{Hartland:2019bjb,Durieux:2019rbz,Brivio:2019ius,correlations1}.

Any evidence of possible LFNU phenomena contradicts the key Standard Model (SM) prediction that
the differences in the rates of processes differing only in the flavor of the leptons involved are suppressed by small differences in the Yukawa couplings. In this sense lepton flavor is an accidental (approximate) symmetry of the SM, which may be strongly violated in a variety of well-motivated NP scenarios. Hence, even if the current experimental indications of LFNU have not yet met discovery criteria, providing an accurate probe of these processes, whether confirming such indications or not, will provide a better understanding of the flavor structure of the physics beyond the SM. 

Let us denote generic lepton(s) + jets processes 
as follows: 
\begin{eqnarray}
(mnp)_{\ell \ell}&:& pp \to \ell_i^+ \ell_i^- + m\cdot j + n \cdot j_b + p \cdot t ~ \label{eq:mnp-channels} \nonumber \\
(mnp)_{\ell} &:& pp \to \ell_i^\pm + m\cdot j + n \cdot j_b + p \cdot t  + \missET ~,
\label{eq:mnp-channels-single}
\end{eqnarray}
where $m$ is the number of light jets ($j$), $n$ is the number of b-jets ($j_b$) and $p$ is the number of top or anti-top quarks in the final state of the leading-order (LO) hard process;
$\missET$ denotes missing transverse energy, associated with final state neutrinos. 
We then define  
two classes of generic LFU tests at the LHC, involving ratios of the charged di-lepton and single-lepton production channels in \eqref{eq:mnp-channels}, normalized to the corresponding electron-production channels:
\begin{eqnarray}
T_{\ell \ell}^{m n p} =
\frac{\sigma_{\ell \ell}^{mnp}}{\sigma_{ee}^{mnp}} ~~~,~~~ 
T_{\ell}^{m n p} =
\frac{\sigma_{\ell}^{mnp}}{\sigma_{e}^{mnp}}
~, \label{eq:R2ell}
\end{eqnarray}
where $\sigma_{\ell \ell}^{mnp}$ and 
$\sigma_{\ell}^{mnp}$ are the total cross-sections 
of the processes $(mnp)_{\ell \ell}$ and 
$(mnp)_{\ell}$ in \eqref{eq:mnp-channels}, respectively. 
Lepton flavor violation effects 
of the type  $pp \to \ell_i \ell_j + \cdots \, (i\not= j)$ will not be considered here. For LFNU processes with only neutrinos in the final state,  ratios such as \eqref{eq:R2ell} are not useful, since the neutrino flavor cannot be detected. In this case a different strategy is needed, which we briefly discuss below. Note that ratio observables such as in \eqref{eq:R2ell} provide more reliable probes of NP, since they potentially minimize the effects of theoretical uncertainties involved in the calculation of the corresponding cross-sections. 
For example, the NLO QCD and, e.g. loop corrections from EFT operators 
(see \cite{Dawson:2018dxp}), are expected to be cancelled to a large extent in our ratio observables, as will all lepton-flavor independent corrections.   
Even so, the impact of the theoretical uncertainties is accounted for in our analysis, as a part of the total systematic uncertainties that we consider below.
Indeed, different variations of ratio observables have been used in recent years for LFNU studies in top-quark decays \cite{Kamenik:2018nxv} and $B$ physics 
\cite{ratioexrev,Aaij:2015yra,Aaij:2017deq,Marzocca,ratioth1,ratioth2,ratioth3,ratioth4,ratioth5,ratioth6,Aaij:2017uff}.

In the SM (or within NP scenarios which conserve lepton flavor universality) we have
$T_{\ell \ell}^{m np},~T_{\ell}^{m np} \to  1$,
since, as noted above, deviations from unity can only be generated  through the non-universal Higgs-lepton Yukawa couplings 
and through lepton mass dependent polynomials and logarithms from higher order
corrections. The former is
proportional to the lepton masses and is therefore
negligible, while the latter are much smaller than the expected
experimental accuracy --  as is the case, {\it in particular}, for high $p_T$ events which is of our interest in this work.
We will include non-universal
reconstruction efficiencies for the different leptonic final states in the overall uncertainty of the measurement of 
$T_{\ell \ell,\,\ell}^{mnp}$ defined in \eqref{eq:R2ell}.

As mentioned earlier, we describe the underlying NP responsible for $T_{\ell \ell}^{mnp} \neq 1$ and $T_{\ell}^{mnp} \neq 1$, using
the SMEFT framework, defined by adding to the SM
Lagrangian  an infinite series of higher-dimensional, gauge-invariant
operators, ${\cal O}_i^{(n)}$. These operators 
are constructed using the SM fields and their coefficients are
suppressed by inverse powers of the NP scale $M$
\cite{EFT1,EFT2,EFT3,EFT4}:
\begin{eqnarray}
{\cal L} = {\cal L}_{SM} + \sum_{n=5}^\infty
\frac{1}{M^{n-4}} \sum_i f_i O_i^{(n)} \label{eq:EFT1}~,
\end{eqnarray}
where $n$ is the mass dimension of $O_i^{(n)}$
and we assume decoupling and weakly-coupled heavy NP, so that
$n$ equals the canonical dimension. The dominating NP effects
are then expected to be generated by
contributing operators
with the lowest dimension (smallest $n$)
that can be generated at tree-level in the underlying theory.  The (Wilson) coefficients $f_i$ depend on the details of the underlying heavy theory and, therefore, parameterize all possible weakly-interacting and decoupling types of heavy physics.

In what follows we will consider the leading dimension six
operators ($n=6$) and drop the index $n$.\footnote{There is a single lepton number violating dimension five operator in the SMEFT framework, the so called Weinberg operator $\bar\ell^c \tilde H^\star
\tilde H^\dagger \ell$ \cite{Weinberg}, for which the scale $\Lambda $ is very large and is therefore not relevant for this study.} We also define the 
``effective scale'' $ \Lambda = M/\sqrt{|f|}$ whence
\begin{equation}
    f/M^2 = \eta_f/\Lambda^2 ~, \label{eff-Lam}
\end{equation}
where $\eta_f = \pm1$ denotes the sign 
of $f$. Thus, for example, 
$\Lambda = M$ for ``natural'' NP with 
$|f| = 1$, which we will assume throughout the rest of this work, unless stated otherwise. 

\begin{table*}[htb]
\caption{\label{tab:LNU-dim6}
The potentially lepton non-universal dimension six operators in the SMEFT (using the Warsaw basis \cite{EFT4}, see also text). The subscripts $p,r,s,t$ are flavor indices.}
\begin{center}
\small
\begin{minipage}[t]{4.45cm}
\renewcommand{\arraystretch}{1.5}
\begin{tabular}[t]{c|c}
\multicolumn{2}{c}{Higgs-Lepton scalar} \\
\hline
${\cal O}_{eH}(pr)$           & $(H^\dag H)(\bar l_p e_r H)$ \\
\end{tabular}
\end{minipage}
\begin{minipage}[t]{4.45cm}
\renewcommand{\arraystretch}{1.5}
\begin{tabular}[t]{c|c}
\multicolumn{2}{c}{Higgs-Lepton vector} \\
\hline
${\cal O}_{H l}^{(1)}(pr)$      & $(H^\dag i\overleftrightarrow{D}_\mu H)(\bar l_p \gamma^\mu l_r)$\\
${\cal O}_{H l}^{(3)}(pr)$      & $(H^\dag i\overleftrightarrow{D}^I_\mu H)(\bar l_p \tau^I \gamma^\mu l_r)$\\
${\cal O}_{H e}(pr)$            & $(H^\dag i\overleftrightarrow{D}_\mu H)(\bar e_p \gamma^\mu e_r)$\\
\end{tabular}
\end{minipage}
\begin{minipage}[t]{4.45cm}
\renewcommand{\arraystretch}{1.5}
\begin{tabular}[t]{c|c}
\multicolumn{2}{c}{Lepton MDM} \\
\hline
${\cal O}_{eW}(pr)$      & $(\bar l_p \sigma^{\mu\nu} e_r) \tau^I H W_{\mu\nu}^I$ \\
${\cal O}_{eB}(pr)$        & $(\bar l_p \sigma^{\mu\nu} e_r) H B_{\mu\nu}$ \\
\end{tabular}
\end{minipage}

\vspace{0.25cm}

\begin{minipage}[t]{4.45cm}
\renewcommand{\arraystretch}{1.5}
\begin{tabular}[t]{c|c}
\multicolumn{2}{c}{$4-{\rm Fermi}:(\bar LL)(\bar LL)$} \\
\hline
${\cal O}_{lq}^{(1)}(prst)$                & $(\bar l_p \gamma_\mu l_r)(\bar q_s \gamma^\mu q_t)$ \\
${\cal O}_{lq}^{(3)}(prst)$                & $(\bar l_p \gamma_\mu \tau^I l_r)(\bar q_s \gamma^\mu \tau^I q_t)$
\end{tabular}
\end{minipage}
\begin{minipage}[t]{4.45cm}
\renewcommand{\arraystretch}{1.5}
\begin{tabular}[t]{c|c}
\multicolumn{2}{c}{$4-{\rm Fermi}:(\bar RR)(\bar RR)$} \\
\hline
${\cal O}_{eu}(prst)$                      & $(\bar e_p \gamma_\mu e_r)(\bar u_s \gamma^\mu u_t)$ \\
${\cal O}_{ed}(prst)$                      & $(\bar e_p \gamma_\mu e_r)(\bar d_s\gamma^\mu d_t)$ \\
\end{tabular}
\end{minipage}
\begin{minipage}[t]{4.45cm}
\renewcommand{\arraystretch}{1.5}
\begin{tabular}[t]{c|c}
\multicolumn{2}{c}{$4-{\rm Fermi}:(\bar LL)(\bar RR)$} \\
\hline
${\cal O}_{lu}(prst)$               & $(\bar l_p \gamma_\mu l_r)(\bar u_s \gamma^\mu u_t)$ \\
${\cal O}_{ld}(prst)$               & $(\bar l_p \gamma_\mu l_r)(\bar d_s \gamma^\mu d_t)$ \\
${\cal O}_{qe}(prst)$               & $(\bar q_p \gamma_\mu q_r)(\bar e_s \gamma^\mu e_t)$ \\
\end{tabular}
\end{minipage}

\vspace{0.25cm}

\begin{minipage}[t]{4.45cm}
\renewcommand{\arraystretch}{1.5}
\begin{tabular}[t]{c|c}
\multicolumn{2}{c}{$4-{\rm Fermi}:(\bar LR)(\bar RL)+\hbox{h.c.}$} \\
\hline
${\cal O}_{ledq}(prst)$ & $(\bar l_p^j e_r)(\bar d_s q_{tj})$
\end{tabular}
\end{minipage}
\begin{minipage}[t]{4.45cm}
\renewcommand{\arraystretch}{1.5}
\begin{tabular}[t]{c|c}
\multicolumn{2}{c}{$4-{\rm Fermi}:(\bar LR)(\bar L R)+\hbox{h.c.}$} \\
\hline
${\cal O}_{lequ}^{(1)}(prst)$ & $(\bar l_p^j e_r) \epsilon_{jk} (\bar q_s^k u_t)$ \\
${\cal O}_{lequ}^{(3)}(prst)$ & $(\bar l_p^j \sigma_{\mu\nu} e_r) \epsilon_{jk} (\bar q_s^k \sigma^{\mu\nu} u_t)$
\end{tabular}
\end{minipage}
\end{center}
\end{table*}

In Table \ref{tab:LNU-dim6} we list all the dimension six operators that can potentially violate LFU and that are, therefore, relevant for this study; operators with four leptons are excluded and we also assume that baryon number is conserved in the underlying heavy theory. 
Here, we will demonstrate 
our strategy for the two specific SU(2) 
triplet and singlet 4-Fermi operators ($prst$ are flavor indices):
\begin{eqnarray}
{\cal O}_{l q}^{(3)}(prst) &=& 
\left(\bar l_p \gamma_\mu \tau^I l_r \right) \left(\bar q_s \gamma^\mu \tau^I q_t \right) 
\label{eq:Olq3}~, \\
{\cal O}_{qe}(prst) &=& 
\left(\bar e_q \gamma_\mu e_r \right)  
\left(\bar q_s \gamma^\mu q_t \right) \label{eq:Oqe}~,
\end{eqnarray}
focusing on the case where the 
heavy underlying NP has 
a LFNU coupling to 3rd generation quarks and
2nd generation leptons, i.e., on ${\cal O}_{l q}^{(3)}(2233)$ and ${\cal O}_{qe}(2233)$; it should be understood, though, that similar effects can be generated in the electron and $\tau$-lepton channels, though, 
the phenomenology and 
detection strategies of final states involving the $\tau$-leptons are fundamentally different from those involving the electrons and muons. Note also that we will adopt a "one-coupling-scheme", i.e, we will study the effects of one operator at the time.
The reasoning behind focusing on one type of NP is that, in general, the scales and dynamics of the NP that underlies the different operators may vary, so that "injecting" into the processes considered below more than one type of NP requires additional assumptions regarding the energy scales and the signs and sizes of the corresponding Wilson Coefficients.

We will not consider operators that have a flavor changing quark current involving the 3rd generation quarks, e.g., ${\cal O}_{l q}^{(3)}(2232)$ and 
${\cal O}_{l q}^{(3)}(3332)$, which 
can generate the $b \to s \mu^+ \mu^-$ and $b \to c \tau^- \bar\nu_\tau$ transitions 
and may, therefore, contribute to 
$R_{K^{(*)}}$ and $R_{D^{(*)}}$, respectively.
These operators can generate LFNU collider signals similar to those studied here, see e.g., \cite{soniRPV,bsll,Altmannshofer:2020axr}.
For example, ${\cal O}_{lq}^{(3)}(2232)$ generates the $\mu^+ \mu^- \bar s b$ and $\mu^+ \mu^- \bar c t$ contact terms with the same effective scale, which can contribute to the ratio observables $T_{\mu \mu}^{010}$ and $T_{\mu \mu}^{001}$, via $sg \to b \mu^+ \mu^- $ and $cg \to t \mu^+ \mu^-$, respectively. Furthermore, the operator ${\cal O}_{lq}^{(3)}(3332)$ generates the contact interactions $\tau^+ \tau^- \bar s b$, $\tau^- \bar \nu_\tau \bar s t$
and $\tau^+ \nu_\tau \bar c b$, which can contribute to the $T$-tests $T_{\tau \tau}^{010}$, $T_{\tau \tau}^{110}$ as well as $T_{\tau}^{010}$, $T_{\tau}^{110}$ via the hard processes $sg \to b \tau^+ \tau^-$, $gg \to \bar s b \tau^+ \tau^-$ and $cg \to \bar b \tau^- \bar\nu_\tau$, $gg \to c \bar b \tau^- \bar\nu_\tau$, respectively.

To study the sensitivity to the flavor non-universal 
NP we define the following $\chi^2$-test:\footnote{\label{foot.2}In the general case, where the correlation matrix for the systematic uncertainties is provided, the $\chi^2$-test 
reads instead (e.g., for the di-muon channels): 
$\chi^2 = \sum_{ij} \left( T_{\ell \ell}^{X_i}(\Lambda) - T_{\ell \ell}^{X_i,exp} \right) \sigma_{X_i X_j}^{-2} \left( T_{\ell \ell}^{X_j}(\Lambda) - T_{\ell \ell}^{X_j,exp} \right)$, where 
$\sigma_{X_i X_j}^{-2} = \left( \delta T^{X_i} \rho^{X_i X_j} \delta T^{X_j} \right)^{-1}$ and $\rho^{X_i X_j}$ is the correlation matrix provided by the experiment (see also discussion below).}  
%
%
%
\begin{eqnarray}
\chi^2 = \sum_{X} \frac{\left[ T_{\ell \ell}^X(\Lambda) - T_{\ell \ell}^{X,\tt exp} \right]^2}{\left(\delta T^{X}\right)^2} +
\sum_{Y} \frac{\left[ T_{\ell}^Y(\Lambda) - T_{\ell}^{Y,\tt exp} \right]^2}{\left(\delta T^{Y}\right)^2}~, \nonumber \\
\label{eq:chi2} 
\end{eqnarray}
where $X,Y \in \left(m,n,p\right)$ denote the $\ell \ell$ and single $\ell$ channels, respectively, and $\delta T^X,\delta T^Y$ denote the corresponding total experimental plus theoretical 
$1 \sigma$ uncertainties, 
which are assumed to be statistically independent (the experimental uncertainties are assumed to be the dominant ones, see also discussion above). 

For the purpose of exacting a bound on $ \Lambda $ we assume that, on average, no NP is observed. We thus generate 
${\cal O}(10000)$ random realizations of the sets of ``measured'' $T$-tests, $T_{\ell \ell}^{X,\tt exp}$ and $T_{\ell}^{Y,\tt exp}$ [to be used for the 
$\chi^2$-test in \eqref{eq:chi2}], normally distributed with average 1 
(i.e., the SM prediction) and standard deviation $ \delta T$:\footnote{Due to the different detection efficiencies of electrons and muons, we expect the ratios $T_{\ell \ell,\ell}^{\tt exp}$ to deviate from unity even in the absence of NP. This, however, has no effect on our $\chi^2$-test analysis and will not change our main results.}
\begin{eqnarray}
T_{\ell \ell}^{X,\tt exp}= {\cal N} \left(1, \left(\delta T^{X} \right)^2 \right), ~ T_{\ell}^{Y,\tt  exp} = {\cal N} \left(1, \left(\delta T^{Y} \right)^2 \right) ~.  \label{eq:normal}  
\end{eqnarray}
where $ {\cal N}(a,s^2)$ denotes the normal distribution for average $a$ and standard deviation $s$.

\begin{figure}[htb]
\includegraphics[width=0.7\textwidth]{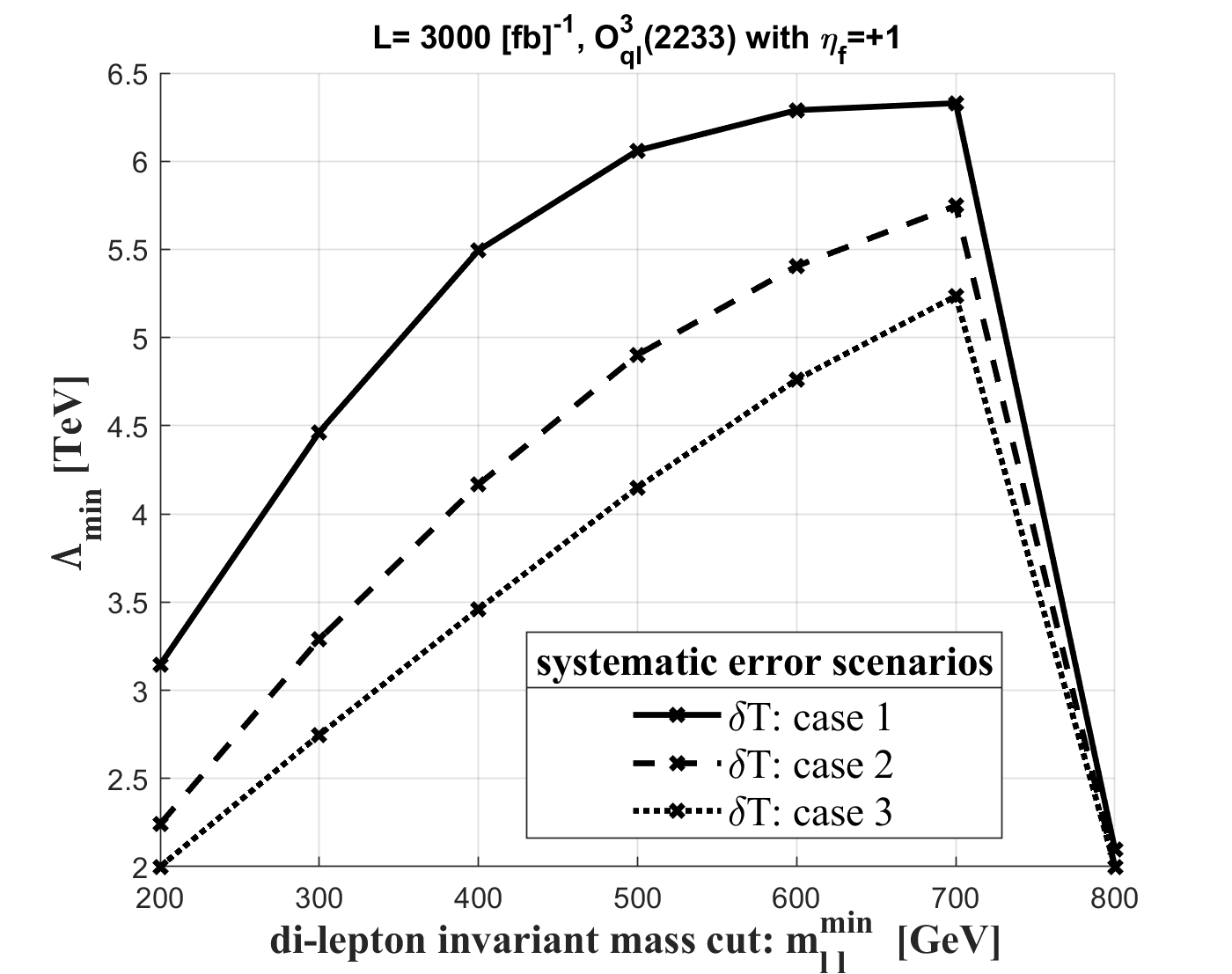}
\caption{Expected 95\% $CL$ bounds on the scale of the SU(2)-triplet operator ${\cal O}_{lq}^{(3)}(2233)$ with $\eta_f=+1$, as a function of the di-lepton invariant mass cut $m_{\ell \ell}^{min}$, for the HL-LHC with an integrated luminosity of $3000$ [fb]$^{-1}$. Results are shown for three overall uncertainty scenarios: $\delta T$ cases 1-3. See also text.}
\label{fig:lumsOlq3}
\end{figure}

The overall uncertainties $\delta T^{X}$ and $\delta T^{Y}$ of the data samples 
 are taken as:
\begin{eqnarray}
\delta T^{X,Y} = 
\sqrt{ \left( \delta T^{X,Y}_{\tt stat} \right)^2 +
\left( \delta T^{X,Y}_{\tt sys}\right)^2}~, \label{eq:error}
\end{eqnarray}
where $\delta T^{X,Y}_{\tt stat}$ and $\delta T^{X,Y}_{\tt sys}$ stand for the statistical and systematic uncertainties expected in the data samples, respectively. The statistical uncertainties are estimated from 
the expected number of events based on the SM cross-sections: $\delta T^{X}_{\tt stat} = \sqrt{2/N_{\ell \ell}^{X}(SM)}$ and $\delta T^{Y}_{\tt stat} = \sqrt{2/N_{\ell}^Y(SM)}$;
for the systematic uncertainties we analyse below 
3 different cases: 
$\delta T^{X,Y}_{\tt sys} = 5\%, 10\%, 15\%$ for channels involving only light-jets 
and/or b-jets in the final state and 
$\delta T^{X,Y}_{\tt sys} = 10\%, 20\%, 30\%$
for channels with a top-quark in the final state.
Without knowing 
the actual uncertainties of the experiment, the uncertainty scenarios outlined above serve as realistic benchmarks for conveying the main message of this work. In particular, we assume that they account for both the experimental and the theoretical uncertainties, while the latter are expected to be minimized due to the use of ratio observables (see also discussion above). Moreover, we assume (in Eq.~\eqref{eq:chi2}) that 
the systematic uncertainties are uncorrelated, since the information about the correlation matrix of the uncertainties is not yet available for the measurements/channels used in our $\chi^2$-test (see also footnote \ref{foot.2}). We note, though, that correlations among the systematic uncertainties in the various channels used below will degrade the sensitivity to the NP, since they effectively reduce the number of observables/channels. For example, a 100\% correlation among the uncertainties of the di-muon channels used below is equivalent to using a single channel, and that can cause a dramatic loss of sensitivity as is shown in Fig.\ref{fig:lumsOlq3}, i.e., comparing the sensitivity to $\Lambda$ with $m_{\ell \ell}^{\rm min} =700$~GeV with that of $\Lambda$ with $m_{\ell \ell}^{\rm min} =800$~GeV.

The expected bounds on $\Lambda$ are then extracted from
the randomly distributed range of best fitted 
values of $\Lambda$ that minimize the $\chi^2$-test of Eq.~\eqref{eq:chi2} (an example is shown in Appendix A).
We use three LHC integrated luminosity scenarios: 
${\cal L}=140,300,3000$ [fb]$^{-1}$, corresponding to the currently accumulated LHC plan, the RUN3 projections and the planned HL-LHC luminosity, respectively.
Then, based on the SM cross-sections, we demand at least 100 events 
for any of the channels $X,Y \in \left(m,n,p\right)$, i.e., $\sigma^{SM} \cdot {\cal L} > 100$, or else this channel is not included in the 
$\chi^2$-test of Eq.~\eqref{eq:chi2}. 
This 100 event criterion is set to ensure that the potential reducible backgrounds (see discussion below) will be sub-leading and, therefore, have a small impact on the overall uncertainty in these measurements.

We demonstrate below our formalism for detecting LFNU 
based on the ratio observables of \eqref{eq:R2ell}, using the QCD generated 
(and therefore dominant)
exclusive di-muon + multi-jet and/or top-quarks channels:\footnote{We do not consider here Drell-Yan di-lepton production $pp \to \ell^+ \ell^-$, i.e., with no jet activity, which correspond to the LFNU signal test $T_{\ell \ell}^{000}$ and which, in our case, are generated by b-quark fusion and are, therefore, sub-leading. Such Drell-Yan processes were studied within the SMEFT framework and in connection to LFNU physics and the B-anomalies in \cite{Marzocca,Admir,Fuentes-Martin:2020lea}, where bounds on the corresponding 4-Fermi operators were derived (see also discussion below).} 
\begin{eqnarray}
(010)_{\mu \mu}&:& pp \to \mu^+ \mu^- + j_b  \nonumber \\
(110)_{\mu \mu}&:& p p \to \mu^+ \mu^- + j+ j_b  \nonumber \\
(020)_{\mu \mu}&:& pp \to \mu^+ \mu^- + 2 \cdot j_b  \nonumber \\
(002)_{\mu \mu}&:& pp \to \mu^+ \mu^- + t \bar t \label{eq:ellell_processes} ~,
\end{eqnarray}
where the LFNU effects are generated by the operators ${\cal O}_{lq}^{(3)}(2233)$
and ${\cal O}_{qe}(2233)$ (i.e.,  the cross-sections involving electrons in the denominator of $T_{\mu \mu}^{mnp}$ in \eqref{eq:R2ell} are assumed to be SM-like).
We can then define a generic form for 
the cross-section in \eqref{eq:ellell_processes} with a cut $m_{\ell \ell} > m_{\ell \ell}^{\tt min}$ on the di-muon invariant mass:
\begin{widetext}
\begin{eqnarray}
\sigma_{\ell \ell}^{mnp}(m_{\ell \ell}^{\tt min})=\sigma_{\ell \ell}^{{\tt SM},mnp}(m_{\ell \ell}^{\tt min}) + \frac{\sigma_{\ell \ell}^{{\tt INT},mnp}(m_{\ell \ell}^{\tt min})}{\Lambda^2} + \frac{\sigma_{\ell \ell}^{{\tt NP},mnp}(m_{\ell \ell}^{\tt min})}{\Lambda^4} ~, \label{EFT_exp}
\end{eqnarray}
\end{widetext}
where $\sigma^{\tt INT}$ and $\sigma^{\tt NP}$ are the SM$\times$NP interference and NP$^2$ terms, respectively. 
The dominant NP contribution then depends on the di-lepton invariant mass cut 
and the di-lepton channel involved. In particular,
the ${\cal O}(\Lambda^{-2})$ correction, $\sigma^{\tt INT}$, dominates for moderate di-lepton invariant mass 
cut, for which the 
SM term is appreciable, whereas 
the ${\cal O}(\Lambda^{-4})$ NP$^2$ correction, $\sigma^{\tt NP}$, is dominant in the high
$m_{\ell \ell}^{\tt min}$-cut regime, where 
the SM contribution is suppressed.
We thus obtain a better sensitivity to the NP with higher 
$m_{\ell \ell}^{\tt min}$-cuts (see below), for which the signal-to-background ratio is significantly improved.


All cross-sections contributing to the LFU $T$-tests in \eqref{eq:R2ell} were calculated exclusively (i.e., separately for each channel without matching) using {\sc MadGraph5\_aMC@NLO} \cite{madgraph5}
at LO parton-level and 
a dedicated universal FeynRules output (UFO) model
for the EFT framework was produced using {\sc FeynRules} \cite{FRpaper}, where we have assumed 
for simplicity that the NP effects reside in the muonic operators, so that the cross-sections involving electrons are SM-like.
In addition, the LO MSTW 2008 parton distribution functions (PDF) set (MSTW2008lo68cl \cite{MSTW2008})\footnote{We note that our results, which are based on ratio observables, are insensitive to the PDF choice (within the uncertainties considered), in particular, 
since the PDF choice is lepton universal and, therefore, has a similar effect on final states with different lepton generations.} in the 5 flavor scheme
was used with a dynamical scale choice for the
central value of the factorization ($\mu_F$)
and renormalization ($\mu_R$) scales, corresponding to the sum of
the transverse mass in the hard-process. 
As a baseline selection, we used the default {\sc MadGraph5\_aMC@NLO} parameters: $p_T > $ 20~GeV and $|\eta| < 5$ for jets, $p_T > $ 10~GeV and $|\eta| < 2.5$ for leptons.
The minimum angular distance in the $\eta-\phi$ plane between all objects (leptons and jets) is $> 0.4$.
Finally,
kinematic selections cuts (e.g., 
on the di-lepton invariant mass) were imposed using {\sc MadAnalysis5} \cite{madanalysis5}.

Before presenting our results, we would like to address the validity of our EFT analysis, in particular, in connection to the high $m_{\ell \ell}^{\tt min}$ regime and the role of the higher-dimensional operators in the EFT expansion of \eqref{EFT_exp}. This has two aspects (see also \cite{bbll}): (i) the validity of the EFT expansion in $1/\Lambda$, i.e., in terms of the scale of the higher dimension operators, and (ii) the validity of the specific cross-section calculations within the EFT prescription. In particular, as mentioned above, the SM$\times$NP interference term is significantly suppressed with the high $m_{\ell \ell}^{\tt min}$ cut that we use and, so, the leading effect comes from the NP$^2$ 
which is $\propto \Lambda^{-4}$. The next term in the EFT expansion would be the SM$\times$NP(dim.8) contribution, where NP(dim.8) stands for dimension eight operators, so that this contribution is also $\propto \Lambda^{-4}$. However, since the SM$\times$NP(dim.8) terms are proportional to the SM amplitude, they are subject to the same suppression at high $m_{\ell \ell}^{\tt min}$ cuts and their contribution is, therefore, expected to be even smaller than the 
sub-leading SM$\times$NP(dim.6) in \eqref{EFT_exp}. In this sense our EFT expansion is valid, even though the NP$^2$ term dominates.  

As for the validity of the calculation within the EFT framework: this is a more subtle issue, since it depends on the details of the underlying theory. Namely, the validity of the EFT calculation naively requires the overall energy flow in the underlying scattering process to be smaller than the NP threshold, i.e., that $\sqrt{\hat s} < \Lambda$, to ensure that the heavy excitations from the underlying NP  
cannot be produced on-shell. In our case, we find that the sensitivity to the NP (i.e., the bounds) reaches $\Lambda \sim 3-6$~TeV for underlying NP couplings of ${\cal O}(1)$, i.e., for a Wilson coefficient $f=1$. Thus, the EFT prescription is valid since the bulk of the generated events are clustered below $\sqrt{\hat s} \sim 3-4$~TeV, due to the energy limitations of the 14~TeV LHC. Also, if the underlying NP couplings correspond to e.g., $f=2$, then our bounds apply to a NP scale of $M = \sqrt{2} \cdot \Lambda$ (see \eqref{eff-Lam}), so that the EFT validity in this case is further expanded to higher c.m. energies. Furthermore, in some instances the EFT approach may hold even if the overall energy flow is larger than the NP scale; for example, if the heavy NP is being exchanged in the t-channel, so that the energy flow through the heavy propagator is effectively lower than $\sqrt{\hat s}$, in which case the EFT prescription also holds when $\sqrt{\hat s} > \Lambda$. Thus, as mentioned above, the consistency of the calculation within the EFT framework as well as the legitimacy of the NP bounds depend on how one interprets the details of the underlying theory \cite{bbll} (see also \cite{Dawson:2018dxp}).

Using the four di-lepton + jets channels in \eqref{eq:ellell_processes}, we show in Fig.~\ref{fig:lumsOlq3} and Table \ref{tab:best-res-Oql3} a sample of the resulting expected 95\% confidence level ($CL$) bounds on scales of the operators ${\cal O}_{lq}^{(3)}(2233)$
and ${\cal O}_{qe}(2233)$,
as a function of the di-muon invariant mass cut $m_{\ell \ell}^{min}$. 
In particular, the Monte Carlo $\chi^2$-test analysis of LFNU was repeated for different values of $m_{\ell \ell}^{\tt min}$, for
the three integrated luminosity cases 
${\cal L}=140,300,3000$ [fb]$^{-1}$ 
and the three systematic uncertainty cases, 
which yield an overall 
uncertainty of 
$\delta T \sim 10\%,15\%,20\%$ for the 
di-muon multi-jets production channels 
$(010)_{\mu \mu}$, $(110)_{\mu \mu}$ and $(020)_{\mu \mu}$, i.e., with no top quarks in the final state. 
The $(002)_{\mu \mu}$ di-muon + top-pair production channel in \eqref{eq:ellell_processes} was not included in the $\chi^2$-test analysis used to derive the 95\% $CL$ bounds listed in Table \ref{tab:best-res-Oql3}, as it does not pass the 100 event criterion for the 
$({\cal L}/[fb^{-1}],m_{\mu \mu}^{\tt min}/[\text{GeV}])=(140,300),(300,400),(3000,700)$ cases considered in this Table. This process, i.e.,  
$pp \to \ell^+ \ell^- + t \bar t$, 
is, however, an important channel that  
might prove to be a promising direction for the future for disentangling various other types of NP effects, e.g., in leptoquark searches \cite{Sirunyan:2018ruf,Bar-Shalom:2018ure}. Note also the sharp drop at
$m_{\mu \mu}^{\tt min} = 800$~GeV, which is caused as a result of our 100 events criteria. In particular, at $m_{\mu \mu}^{\tt min} = 800$~GeV only the $(020)_{\mu \mu}$ channel produces more than 100 events (hence the much lower sensitivity to the NP scale), whereas for
$m_{\mu \mu}^{\tt min} \leq 700$~GeV all three channels with no top-quarks in the final state, i.e., $(010)_{\mu \mu}$, $(110)_{\mu \mu}$ and $(020)_{\mu \mu}$, pass the 100 events criteria.

%

We see that, as expected, the sensitivity to the underlying NP depends 
on the sign of the Wilson coefficients $\eta_f=\pm 1$ and on the overall uncertainty,
and it varies with the 
di-muon invariant mass cut. 
We find, for example, that with the current LHC accumulated luminosity of ${\cal L}=140$ [fb]$^{-1}$, the best 95\% $CL$ bounds are obtained with the cut
$m_{\mu \mu}^{\tt min} = 300$~GeV: $\Lambda \gsim 2.3 - 3.4$~TeV for ${\cal O}_{lq}^{(3)}(2233)$ and $\Lambda \gsim 2.4 - 4.2$~TeV for ${\cal O}_{q e}(2233)$, depending on the 
overall systematic uncertainty and on the sign of $\eta_f$. Also, a much higher sensitivity is expected at the HL-LHC with a tighter cut of $m_{\ell \ell}^{\tt min} = 700$~GeV, reaching up to
$\Lambda \gsim 6.5$~TeV for ${\cal O}_{lq}^{(3)}(2233)$ with $\eta_f=+1$ and 
${\cal O}_{q e}(2233)$ with $\eta_f=-1$. 

\begin{table*}[htb]
\caption{\label{tab:best-res-Oql3} The expected 95\% $CL$ bound on the scale (in TeV) of the 
operators $\mathcal{O}_{lq}^{(3)}(2233)$ and 
$\mathcal{O}_{qe}(2233)$ (in parenthesis), 
for di-muon invariant mass cuts $m_{\mu \mu} > 300, 400$ and $700$~GeV which are applied for an integrated luminosity of 
${\cal L}=140,300$ and $3000$ [fb]$^{-1}$, respectively.  
Results are shown for $\eta_f = \pm 1$ and three values of the overall uncertainty 
of $\delta T \sim 10\%, 15\%$ and $20\%$,  corresponding
to the three systematic uncertainty cases 1,2, and 3.
For all cases considered in the table only the channels 
$(010)_{\mu \mu}$, $(110)_{\mu \mu}$ and $(020)_{\mu \mu}$ pass the 100 criteria. See also text.}
\begin{tabular}{c|c|c|c|c|c|c|}
\multirow{2}{*}{} \
 & \multicolumn{6}{c|}{95\% $CL$ bounds: $\Lambda_{\mathcal{O}_{lq}^{\left(3\right)}(2233)}\left(\Lambda_{\mathcal{O}_{qe}(2233)}\right)$ [TeV]} 
\tabularnewline
\cline{2-7} 
\multirow{2}{*}{} \
& \multicolumn{2}{c|}{${\cal L}=140$ [fb]$^{-1}$} &
 \multicolumn{2}{c|}{${\cal L}=300$ [fb]$^{-1}$} &
 \multicolumn{2}{c|}{${\cal L}=3000$ [fb]$^{-1}$} 
 \tabularnewline
 & \multicolumn{2}{c|}{$m_{\mu \mu}^{\tt min} = 300$~GeV} & 
 \multicolumn{2}{c|}{$m_{\mu \mu}^{\tt min} = 400$~GeV} &
 \multicolumn{2}{c|}{$m_{\mu \mu}^{\tt min} = 700$~GeV}
\tabularnewline
\cline{2-7} 
& $\eta_f=+1$ & $\eta_f=-1$ & $\eta_f=+1$ & $\eta_f=-1$ & $\eta_f=+1$ & $\eta_f=-1$ 
\tabularnewline
\hline 
\
$\delta T \sim 10\%$ (case 1)
 & 3.4(2.6) & 3.2(4.2) & 4.1(3.1) & 3.9(4.9)& 6.3(4.4) & 6.0(6.4)  \tabularnewline
\hline 
\
$\delta T \sim 15\%$ (case 2) & 3.0(2.5) & 2.7(3.8) & 3.7(3.0) & 3.3(4.5) & 5.8(4.2) & 5.5(5.9)  \tabularnewline
\hline 
\
$\delta T \sim 20\%$ (case 3) & 2.6(2.4) & 2.3(3.4) & 3.3(2.9) & 2.9(4.2) & 5.1(4.1) & 4.7(5.5)  \tabularnewline
\hline 
\end{tabular}
\end{table*}

We now consider a complementary analysis where,
instead of 
examining the bounds under the assumption of no NP in the data, we ask what is the discovery potential of a given NP scenario if the NP is assumed to be present in the data. We thus assume that the experimentally measured ratios $T_{\ell \ell}^{X,\tt exp}$ 
are controlled by the NP, so that, in this case, they are normally distributed with a mean value corresponding to the NP expectations 
$T(\bar\Lambda)$:
\begin{eqnarray}
T_{\ell \ell}^{X,\tt exp} &=& {\cal N} \left(T_{\ell \ell}^X(\bar\Lambda), \left(\delta T^{X} \right)^2 \right) ~, \label{eq:normal2}  
\end{eqnarray}
where here $\bar\Lambda$ is the value of the NP scale 
injected into the data and tested against the SM 
prediction. 
As for the overall uncertainties, $\delta T^{X}$, we follow the prescription of~\eqref{eq:error}, where this time the statistical uncertainties are assumed to reflect the NP data, i.e, 
$\delta T^{X}_{\tt stat} = \sqrt{2/N^{X}(\bar\Lambda)}$, where $N^{X}(\bar\Lambda)$ are the number of events expected for a NP scale $\bar\Lambda$ in each of the di-lepton channels $X \in (mnp)$. The systematic uncertainties,
$\delta T^{X}_{\tt sys}$, are kept unchanged, i.e., using the three cases $\delta T^{X}_{\tt sys} = 5\%, 10\%, 15\%$ for channels involving only light-jets 
and/or b-jets in the final state. 
We thus vary $\Lambda$ in the $\chi^2$-test 
of \eqref{eq:chi2} [i.e., with $T^{\tt exp}$ normally distributed around $T(\bar\Lambda)$ following
\eqref{eq:normal2}], from which we 
generate the distribution of the best fitted NP scale, $\hat\Lambda$, for each value of 
 $\bar\Lambda$ (an example is shown in Appendix A).\footnote{In an ideal measurement with $\delta T \to 0$ we will clearly have
$\hat\Lambda = \bar\Lambda$, so that the NP signal corresponding to any $\Lambda$ will be well separated from the SM prediction.} 
This is repeated for different values of $\bar\Lambda$ until we find the value
that yields a distribution 
which deviates from the SM prediction at a given $CL$; we denote this value by $\bar\Lambda(CL)$. 
In Table \ref{tab:best-res-disc} we list a 
sample of our results for the discovery potential of the operators ${\cal O}_{lq}^{(3)}(2233)$ and ${\cal O}_{qe}(2233)$ at the LHC. 
In particular, we find that a 
$5\sigma$ discovery of the 
heavy underlying NP that generates 
these operators can be obtained at 
the LHC with ${\cal L}=300$~fb$^{-1}$, 
if its 
scale is in the range $\bar\Lambda(5 \sigma) \sim 2.3-2.9$~TeV for ${\cal O}_{lq}^{(3)}(2233)$ 
and $\bar\Lambda(5 \sigma) \sim 2.8-3.4$~TeV for ${\cal O}_{qe}(2233)$, depending on the uncertainty 
in the measurement of the ratios $T_{\ell \ell}^{mnp}$. 
At the HL-HLC, the corresponding discovery potential is extended up to $\bar\Lambda(5 \sigma) \sim 3.7 - 4.6$~TeV. 

\begin{table}[htb]
\caption{\label{tab:best-res-disc} The values of the NP scale $\bar\Lambda(CL)$ (in TeV)
that will yield a $5 \sigma$ discovery of the the operators ${\cal O}_{lq}^{(3)}$ with $\eta_f = + 1$ and ${\cal O}_{qe}$ with $\eta_f = - 1$, at the LHC with ${\cal L}=300$~fb$^{-1}$ and $m_{\mu \mu}^{\tt min} = 400$~GeV and at the HL-LHC with
$3000$~fb$^{-1}$ and $m_{\mu \mu}^{\tt min} = 700$~GeV. 
Numbers are given for the three different
overall uncertainties corresponding to cases 1,2,3 of the systematic uncertainties and the channels that pass the 100 criteria for all cases are
$(010)_{\mu \mu}$, $(110)_{\mu \mu}$ and $(020)_{\mu \mu}$. See also text.}
\begin{center}
\begin{tabular}{c|c|c|c|c|}
\multirow{2}{*}{} \
 & \multicolumn{4}{c|}{$5\sigma$ discovery: $\bar\Lambda(5\sigma)$ [TeV]} 
\tabularnewline
\cline{2-5} 
\multirow{2}{*}{} \
 & \multicolumn{2}{c|}{${\cal O}_{lq}^{(3)} (\eta_f=+1)$}& \multicolumn{2}{c|}{${\cal O}_{qe} (\eta_f=-1)$}
 \tabularnewline
\cline{2-5} 
 & $300$~fb$^{-1}$ & $3000$~fb$^{-1}$ & $300$~fb$^{-1}$ & $3000$~fb$^{-1}$ \tabularnewline
\hline 
\
$\delta T$ case 1
 & 2.9 & 4.6 & 3.4 & 4.6  \tabularnewline
\hline 
\
$\delta T$ case 2
 & 2.4 & 4.1 & 3.1 & 4.3 \tabularnewline
\hline 
\
$\delta T$ case 3
 &2.3 & 3.7 & 2.8 & 4.1  
 \tabularnewline
\hline 
\end{tabular}
\par\end{center}
\end{table}

Let us briefly address the potential background for the multi-jets $\ell^+ \ell^-$ production channels $(010)_{\mu \mu}$, $(110)_{\mu \mu}$ and $(020)_{\mu \mu}$ used in our $\chi^2$-tests.  
We note that the irreducible background to these processes such as $Z+jets$ and $W+jets$  production, are  included in our calculation since they interfere with our signals. 
As for the reducible background, the dominant ones 
are single top + W-boson ($tW$: $pp \to t W$) and vector-boson pair production ($VV$: $pp \to VV$) for the $(010)_{\mu \mu}$ and $(110)_{\mu \mu}$ channels (i.e., for the channels $pp \to \mu^+ \mu^- +j$ and $pp \to \mu^+ \mu^- +j + j_b$). For the  
$(020)_{\mu \mu}$ channel, $pp \to \mu^+ \mu^- + 2 \cdot j_b$, 
the $tW$ and $VV$ background 
are sub-leading and the dominant background comes from the more challenging top-quark pair production ($t\bar{t}$: $pp \to t \bar t$). 
Note, however, that as opposed to our leading di-lepton signals $(010)_{\mu \mu}$, $(110)_{\mu \mu}$ and $(020)_{\mu \mu}$, the
reducible background processes, $tW,~VV$ and $t\bar{t}$, involve large missing energy, which is carried by the neutrinos in the final state.
Thus, they can be significantly  
suppressed with a proper selection cut on the missing transverse energy $\missET$ (see also next paragraph) 
and other acceptance criteria such as 
lepton isolation criteria (that can be applied to minimize to the few percent level the contamination from fake non-prompt leptons from either a misidentified hadron or a decay product of a heavy or light flavor hadron, see e.g., \cite{Aaboud:2017buh}) as well as properties of the transverse momenta 
and energy distribution of the final state particles which can be used, e.g., for 
a better separation of the $t\bar{t}$ background from the di-lepton + jets NP signal,
see e.g., \cite{bbll,Sirunyan:2017yrk,CMS:2019see,Tornambe:2018ulr}.

To give a flavor of the signal ($S$) to background ($B$) handle for our LFNU processes, 
we have applied the di-lepton invariant mass cuts that we used above for our signals
$S=(010)_{\mu \mu},~(110)_{\mu \mu}$ and $(020)_{\mu \mu}$ (i.e., $m_{\ell \ell}^{\rm min} = 300, 400, 700$~GeV for ${\cal L}=140,300,3000$~fb$^{-1}$, respectively) and an additional simple 
pre-selection cut of $\missET < 50$~GeV to the leading background processes 
mentioned above $B=tW, ~VV$ and $t\bar{t}$ (the missing energy pre-selection have a negligible effect on our signals). We then obtain 
$B_{(010)_{\mu \mu}} \approx 33,~7.5,~0.5$ [fb], 
$B_{(110)_{\mu \mu}} \approx 10.3,~3.5,~0.2$ [fb] and 
$B_{(020)_{\mu \mu}} \approx 319,~106,~7.8$ [fb] 
for the reducible background to the signal channels $(010)_{\mu \mu},~(110)_{\mu \mu}$ and $(020)_{\mu \mu}$, respectively, where the three values are for $m_{\ell \ell}^{\rm min} = 300, 400, 700$~GeV with ${\cal L}=140,300,3000$~fb$^{-1}$, respectively.    
Thus, using  $N_{SD}=S/\sqrt{B + (\sigma_B \cdot B)^2}$ as a signal-to-background sensitivity "measure", where $\sigma_B \cdot B$ stands for the expected systematic uncertainty, and setting $\sigma_B=10\%$, we obtain for all three integrated luminosity cases 
considered above $N_{SD} \gsim {\cal O}(1)$ for the signal channels 
$(010)_{\mu \mu},~(110)_{\mu \mu}$, whereas $N_{SD} \gsim {\cal O}(0.1)$ for
the more challenging $(020)_{\mu \mu}$ channel.
Note that the $t\bar{t}$ background 
also applies to the $(110)_{\mu \mu}$ channel due to non-ideal b-tagging efficiencies, and although it can be dramatically reduced with the dilepton invariant mass cuts and the missing energy pre-selection $\missET < 50$~GeV, it is still 
more challenging and, as mentioned above, it requires a more elaborated study which is beyond the scope of this paper; see for example the study of \cite{bbll}, which considered the effects of the $\ell^+ \ell^- \bar b b$ contact terms on the $(020)_{\mu \mu}$ channel and performed a detailed signal to background optimization, specifically including the $t\bar{t}$ background.

\begin{figure}[htb]
\centering
\includegraphics[width=0.7\textwidth]{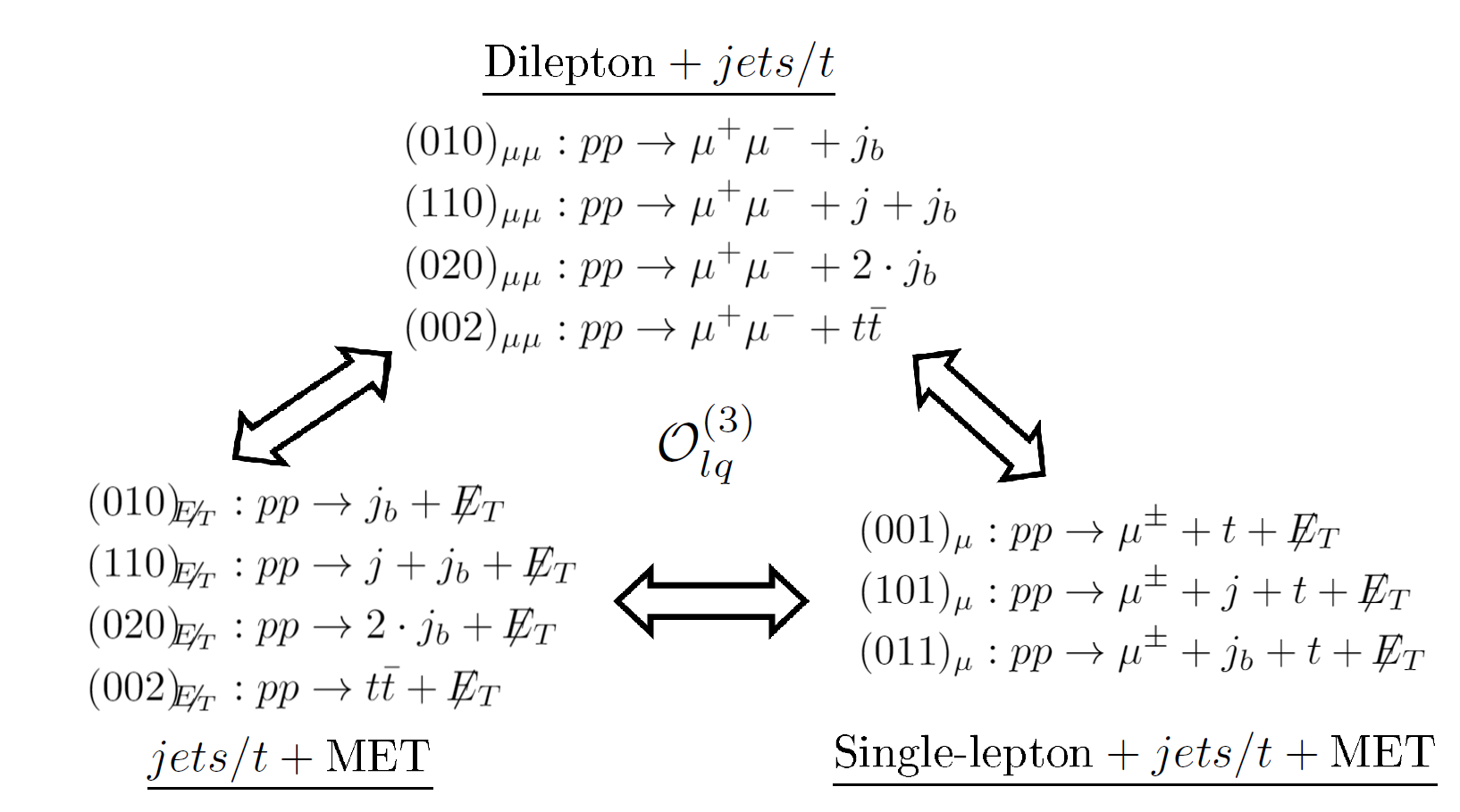}
\caption{Correlations among various di-lepton + jets/tops, single-lepton + jets/tops + $\missET$ and jets/tops + $\missET$ (with no charged lepton) processes, which are generated by the SU(2)-triplet-exchange operator ${\cal O}_{\mu q_3}^{(3)} = (\bar\mu \gamma_\mu \tau^I \mu)(\bar q_3 \gamma^\mu \tau^I q_3)$ via QCD interactions. See also text. }
\centering
\label{fig:correlation}
\end{figure}

Finally, we note that an important difference between the two operators ${\cal O}_{lq}^{(3)}$ and ${\cal O}_{qe}$ with respect to our $\chi^2$-test, 
is that the former also gives rise to 
the single-muon + jets and top-quarks production channels $(mnp)_\mu$ [{\it cf.} \eqref{eq:mnp-channels-single}]
and to neutrino pair-production (with $\missET$ signature) in association with jets and top-quarks: $(mnp)_{\,\missET}: pp \to m \cdot j + n \cdot j_b + p \cdot t + \missET$. 
In particular, these 
$(mnp)_{\mu}$ and $(mnp)_{\,\missET}$ processes are correlated with the di-lepton + jets production channels $(mnp)_{\mu\mu}$ discussed above. For this case also
the best sensitivity to ${\cal O}_{lq}^{(3)}(2233)$ 
is expected from the processes generated by QCD interactions, which are depicted in Fig.~\ref{fig:correlation}.
In particular, note that the QCD-generated single-muon production channels involve a single top-quark in the final state: $(001)_{\mu}$, $(101)_{\mu}$ and $(011)_{\mu}$, 
which affect the ratios $T_{\mu}^{001}$, $T_{\mu}^{101}$ and $T_{\mu}^{011}$ in \eqref{eq:R2ell}. Indeed, we find that including these single-muon + top-quark channels in the $\chi^2$-test of \eqref{eq:chi2} yields a better sensitivity to the scale of this operator, e.g., for the HL-LHC case, yielding bounds which are $\sim 1$~TeV stronger than the ones given in Table \ref{tab:best-res-Oql3}. 
We Note that searches for these single-lepton signatures are on-going (see e.g., \cite{ATLAS:2020llc}), 
since they may be important for searches of various different types of NP, e.g., of pair-production of a scalar partner of the top-quark in supersymmetric theories \cite{Aaboud:2017aeu}. 

Furthermore, the dominant (QCD generated) neutrino channels are the processes $(010)_{\,\missET}$, $(110)_{\,\missET}$, $(020)_{\,\missET}$ and 
$(002)_{\,\missET}$ (see Fig.~\ref{fig:correlation}), which can be used as well to obtain a better sensitivity to this operator. This requires, however, a different approach (rather than our $\chi^2$-tests based on ratio observables) for disentangling the NP  effects and will, therefore, not be further investigated here. 
Note, though, that some of these $\missET$ + jets and/or top-quarks signals are 
important signals of other well motivated NP scenarios. For example,
the processes  $(020)_{\,\missET}$ and 
$(002)_{\,\missET}$, i.e., 
pair production of top-quarks and/or b-jets in association with large $\missET$,
are also signatures of leptoquark pair-production (see e.g., \cite{Bar-Shalom:2018ure,Aaboud:2019jcc}), of
pair-production of the scalar partners of the top or bottom quarks in supersymmetric theories (see e.g., \cite{Aaboud:2017ayj}) and may also be useful for dark matter searches (see e.g., \cite{Aaboud:2017rzf,Sirunyan:2019gfm}).

The approach described above improves on the results obtained in previous interesting studies which are based on the analysis of a single process. For example \cite{Marzocca} obtains limits of $\Lambda > 1.5 - 1.8$~TeV ($\Lambda > 2.5 - 3$~TeV) at the current LHC (HL-LHC) for the scale of the operators 
${\cal O}_{l q}^{(3)}(2233)$ and ${\cal O}_{qe}(2233)$ in \eqref{eq:Olq3} and \eqref{eq:Oqe}, using Drell-Yan di-lepton production $q \bar q \to \ell^+\ell^-$; they find a slight improvement for 4-Fermi operators of 
type $ {\cal O}(1133)$. Note that the current best bound for this last type of operators, i.e., ${\cal O}(1133)$, 
was obtained at LEP  \cite{Ackerstaff:1997nf,Abbiendi:1998ea,Barate:1999qx,Schael:2006wu}: $\Lambda > 0.7 - 2.7$~TeV. We also note that bounds on ${\cal O}_{lq}^{(3)}$ derived from the top-quark decays are significantly weaker \cite{topdecay1,topdecay2} than ours.


To summarize, we have shown
that the 
lepton flavor non-universal ratio observables $T_{\ell \ell}^{m np }$ and $T_{\ell}^{m np }$ of \eqref{eq:R2ell} can be used to 
search for new physics using a $\chi^2$ test that is sensitive to the correlations among several lepton + jets and top-quark production channels.
We found, for example, that with a realistic assessment of the expected uncertainties involved, a 95\%CL bound of 
$\Lambda \gsim 3-4$~TeV can be obtained with the current LHC luminosity, while $\Lambda \gsim 6-7$~TeV is expected at the HL-LHC, for the 4-Fermi operators
${\cal O}_{lq}^{(3)}(2233)$ and 
${\cal O}_{q e}(2233)$ in \eqref{eq:Olq3} and \eqref{eq:Oqe}, 
which involve 2nd generation leptons 
and 3rd generation quarks.
These bounds are obtained with 
a generic di-lepton invariant mass cut for all channels, i.e., without any channel-dependent (specific) optimizations
that, we believe, can
be further used to better isolate the NP effects and, therefore, to obtain an enhanced 
sensitivity to its scale. Though the above discussion involves only 3rd generation quarks, our multi-channel correlation LFU tests are expected to yield an improved sensitivity also for lepton flavor non-universality new physics which involves the 1st and 2nd quark generations.




\acknowledgments
We thank Yoram Rozen for useful discussions. The work of AS  was supported in part by the U.S. DOE contract \#DE-SC0012704.

\bibliographystyle{hunsrt.bst}
\bibliography{mybib2}

\newpage

\appendix


\section{Distributions: bounds and discovery}

\begin{figure*}[htb]
\centering
\includegraphics[width=0.32\textwidth]{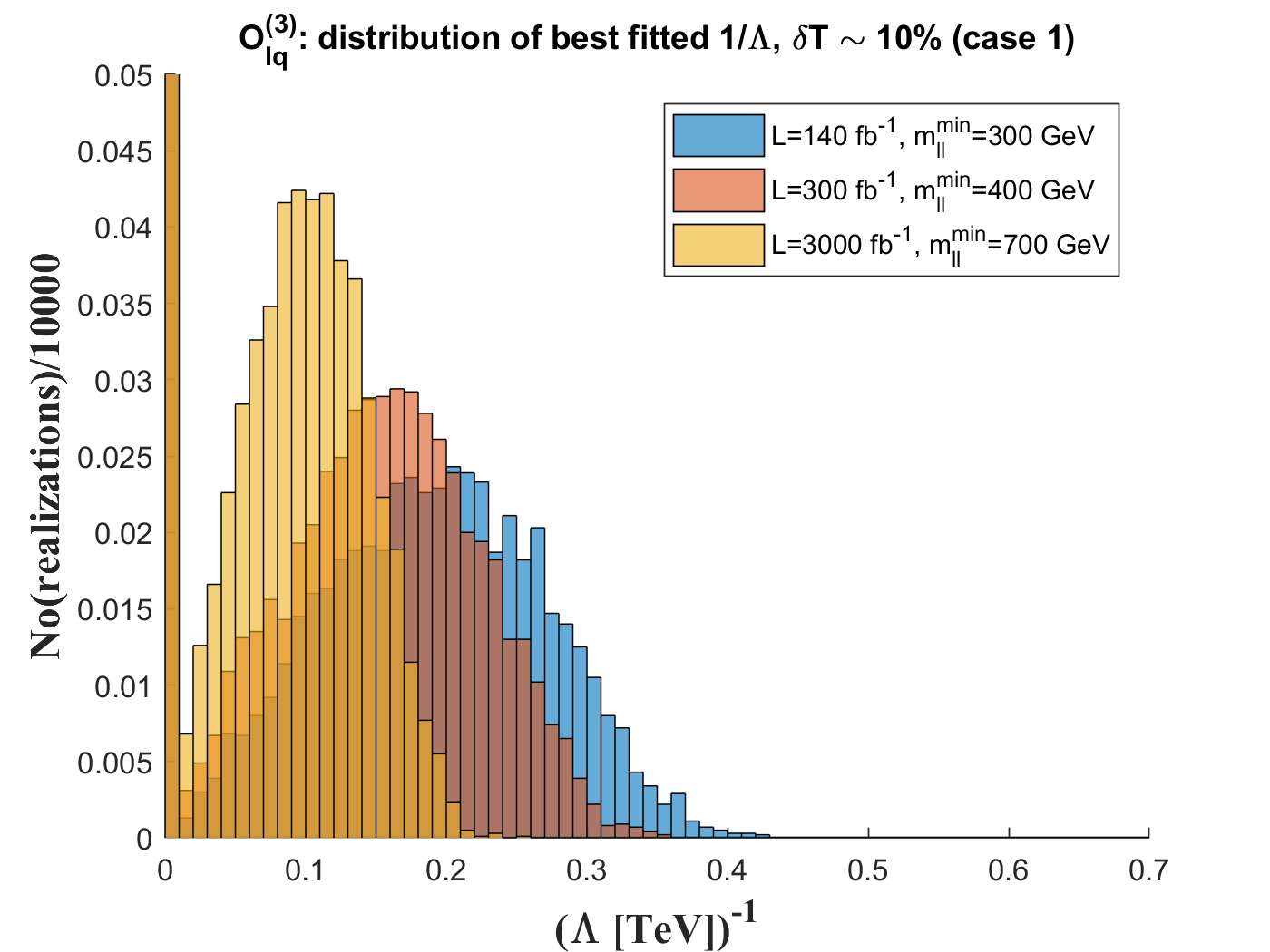}
\includegraphics[width=0.32\textwidth]{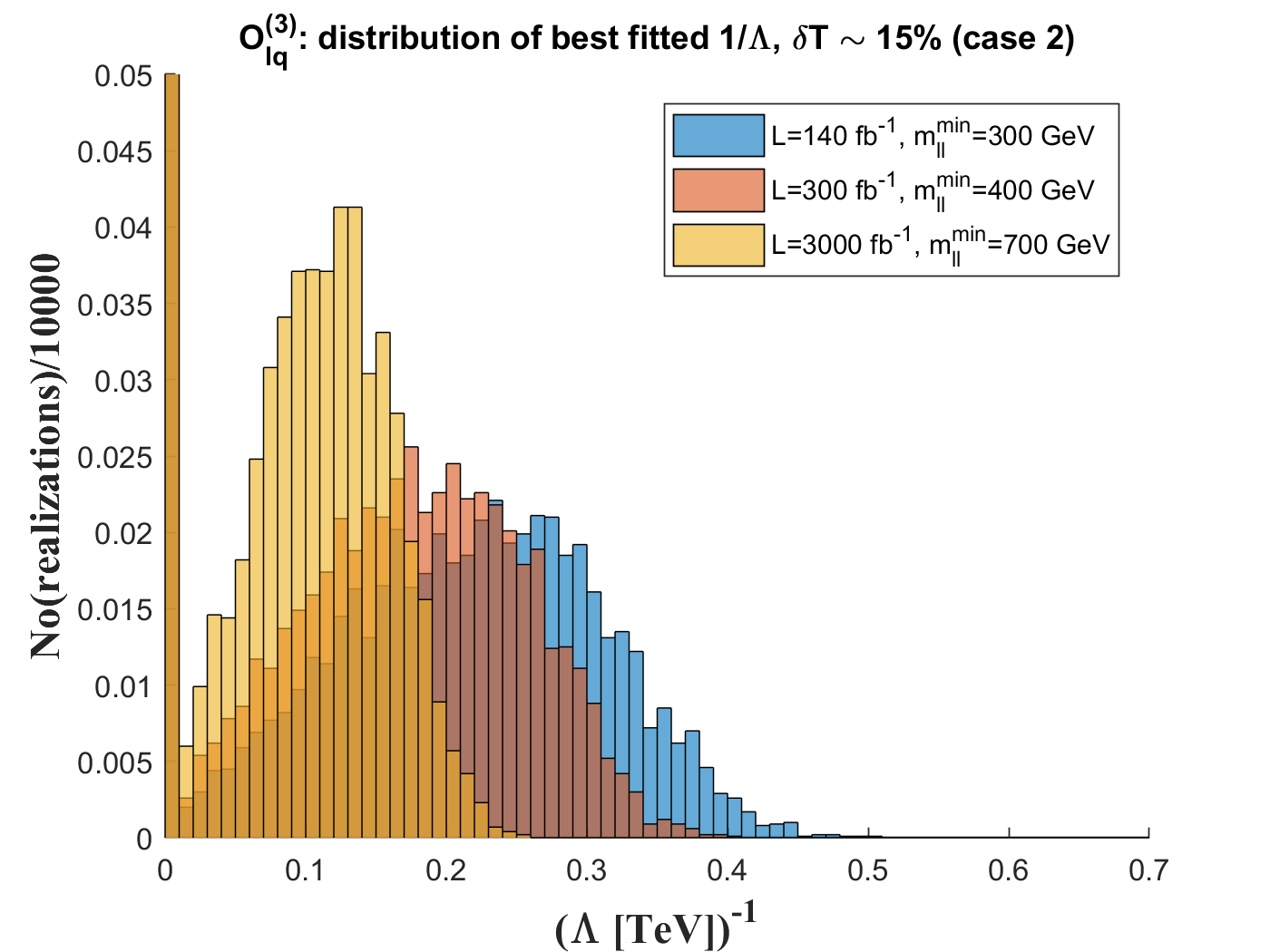}
\includegraphics[width=0.32\textwidth]{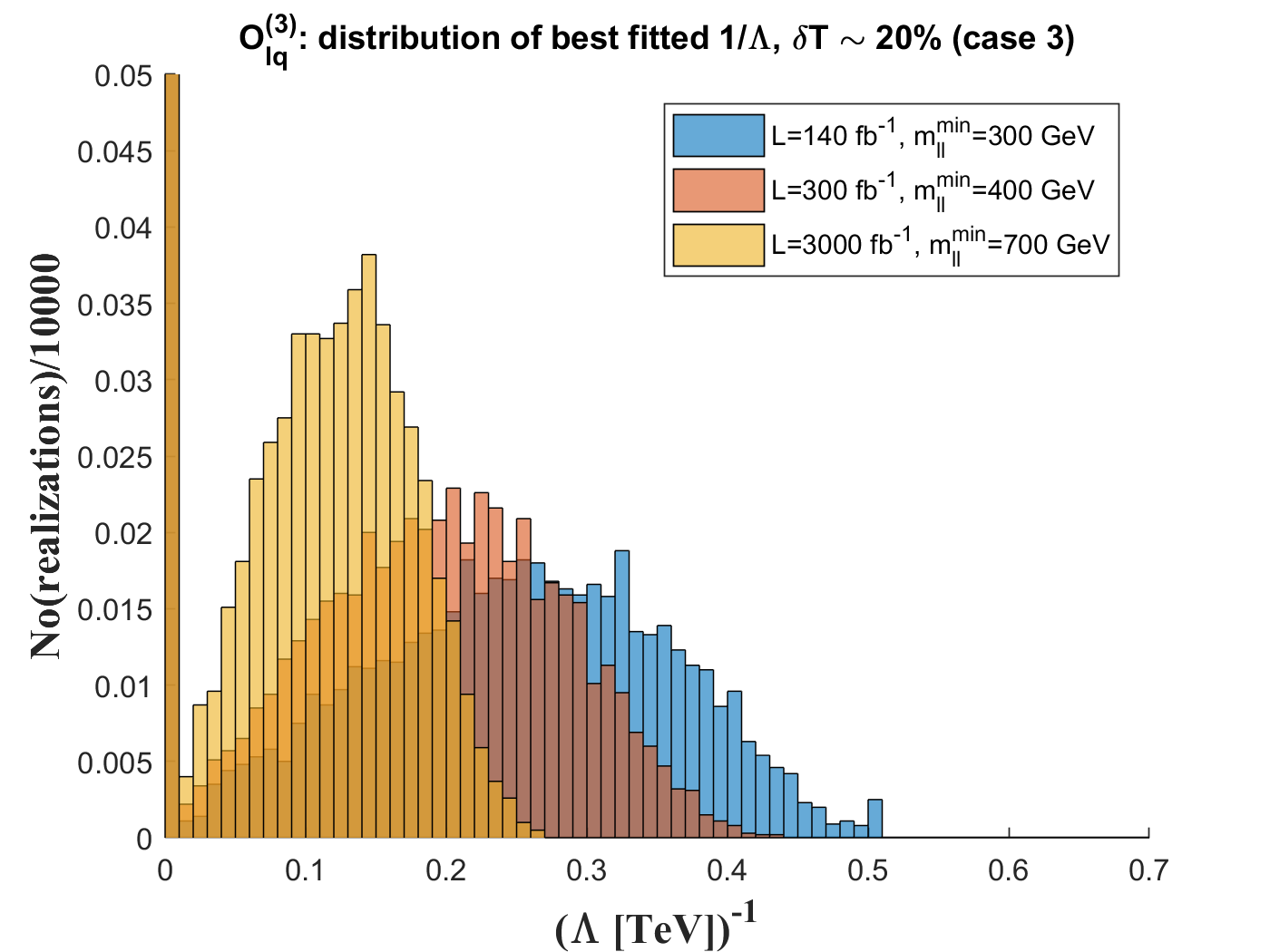}
\caption{The normalized distribution of the inverse value of the best fitted $\Lambda$ of the operator ${\cal O}_{\mu q_3}^{(3)} = (\bar\mu \gamma_\mu \tau^I \mu)(\bar q_3 \gamma^\mu \tau^I q_3)$, that minimize the $\chi^2$-test with $T_{\ell \ell}^{exp}=1$, i.e., corresponding to the SM value, and normally distributed with the three uncertainty scenarios: $\delta T=10\%$ (left), $\delta T=15\%$ (middle) and $\delta T=20\%$ (right). See also text.}
\centering
\label{fig:dist1}
\end{figure*}

In Fig.~\ref{fig:dist1}
we plot the distributions of the best fitted values of $1/\Lambda$ for the operator ${\cal O}_{\mu q_3}^{(3)}(2233)$, that minimize the $\chi^2$-test by mimicking a realistic setting with ${\cal O}(10000)$ random realizations of the experimental values for the LFNU ratios $T_{\ell \ell}^{mnp}=1$, i.e., corresponding to the SM value, and normally distributed with the three uncertainty scenarios: $\delta T=10\%,15\%,20\%$ which correspond to the three 
systematic uncertainty choices outlined in the paper.   
The distributions are shown for 
${\cal L}=140$~fb$^{-1}$ and the selection $m_{\ell \ell}^{min} =300$~GeV, ${\cal L}=300$~fb$^{-1}$ and the selection $m_{\ell \ell}^{min} =400$~GeV and ${\cal L}=3000$~fb$^{-1}$ with the selection $m_{\ell \ell}^{min} =700$~GeV.
The 95\%CL bounds on $\Lambda$ are then extracted from these distributions.  

\begin{figure*}[htb]
\centering
\includegraphics[width=0.35\textwidth]{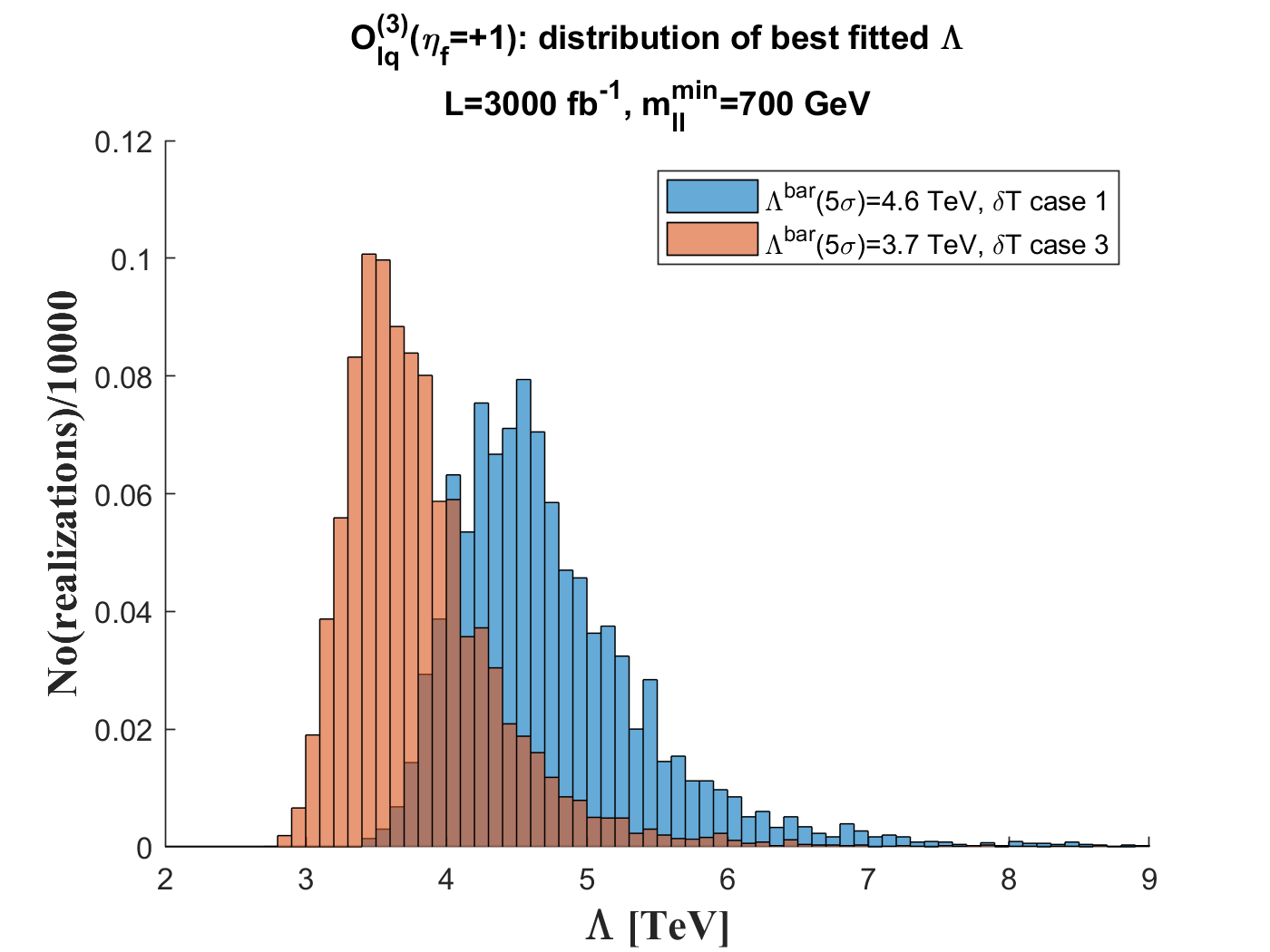}
\includegraphics[width=0.35\textwidth]{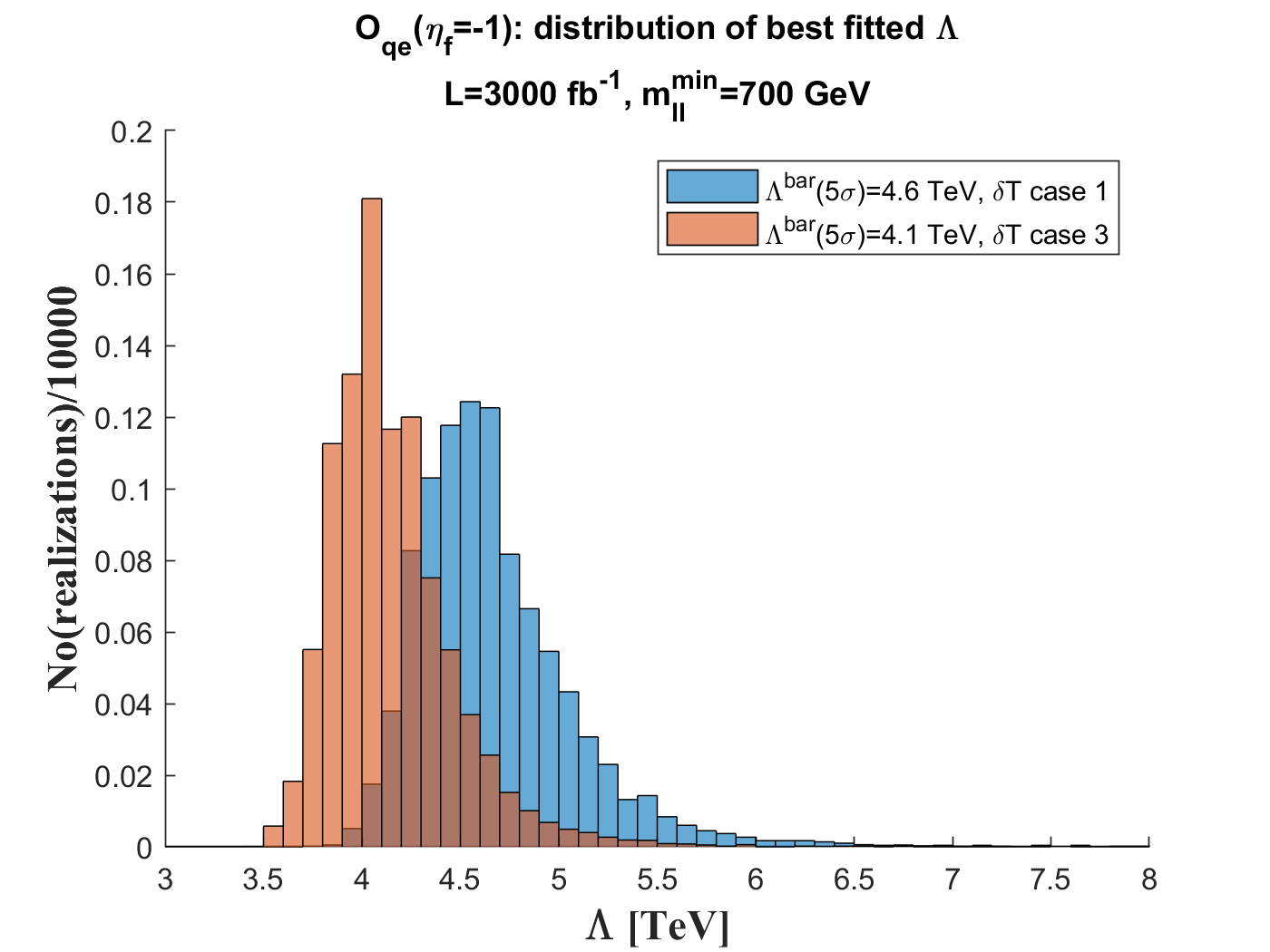}
\caption{The normalized distribution of the best fitted $\Lambda$ 
of the operators ${\cal O}_{l  q_3}^{(3)}(2233)$ (left) and ${\cal O}_{qe}(2233)$ (right), that minimize the $\chi^2$-test with $T_{\ell \ell}^{exp}=T_{\ell \ell}(\bar\Lambda)$, i.e., corresponding to the case where the NP is assumed in the data with specific values of $\bar\Lambda$ (as indicated) and normally distributed with two uncertainty scenarios: $\delta T=10\%$ (case 1) and $\delta T=20\%$ (case 3). See also text.}
\centering
\label{fig:dist2}
\end{figure*}

In Figs.~\ref{fig:dist2}
we plot the distributions of the best fitted values of $\Lambda$ for both the operators ${\cal O}_{\ell q}^{(3)}(2233)$ and ${\cal O}_{qe}(2233)$, that minimize the $\chi^2$-test with ${\cal O}(10000)$ random realizations of the experimental measured ratios $T_{\ell \ell}^{exp}$, corresponding to the case where the NP is assumed in the data with specific values of $\bar\Lambda$, i.e., $T_{\ell \ell}^{exp}=T_{\ell \ell}(\bar\Lambda)$, and normally distributed with two uncertainty scenarios: $\delta T=10\%$ (case 1) and $\delta T=20\%$ (case 3). That is, the experimental values $T_{\ell \ell}^{exp}$ are simulated ${\cal O}(10000)$ times from the normal 
distribution: 
\begin{eqnarray}
T_{\ell \ell}^{X,exp} &=& {\cal N} \left(T_{\ell \ell}^X(\bar\Lambda), \left(\delta T^{X} \right)^2 \right) ~, 
\end{eqnarray}
where $X \in (mnp)$ denotes the di-lepton + jets  production channels, and for each realization we find the best fitted value of $\Lambda$. 

The distributions are shown In Figs.~\ref{fig:dist2} for values of $\bar\Lambda$ that can be discovered at $5\sigma$ at the HL-LHC with ${\cal L}=3000$~fb$^{-1}$ and with the selection of $m_{\ell\ell}^{min} =700$~GeV.

\end{document}